\documentclass[12pt]{iopart}
\usepackage{iopams}
\usepackage{graphicx}
\usepackage{cite}

\begin{document}

\title[Photoemission from the gas phase: An investigation of the space-charge effects]{Photoemission from the gas phase using soft x-ray fs pulses: An investigation of the space-charge effects}

\author{Adriano Verna$^1$, Giovanni Stefani$^1$, Francesco Offi$^1$, Tatsuo~Gejo$^2$, Yoshihito~Tanaka$^2$, Kenta~Tanaka$^2$, Tatsuru~Nishie$^2$, Kiyonobu~Nagaya$^3$, Akinobu~Niozu$^3$, Ryosuke~Yamamura$^4$, Taiga~Suenaga$^4$, Osamu~Takahashi$^4$, Hikaru~Fujise$^5$\footnote{Present address: Department of Chemistry, Graduate School of Science, Nagoya University, Chikusa-ku, Nagoya, Aichi 464-8602, Japan}, Tadashi~Togashi$^6$, Makina~Yabashi$^7$ and Masaki~Oura$^7$}

\address{$^1$ Dipartimento di Scienze, Universit\`a degli Studi Roma Tre, Via della Vasca Navale 84, 00146 Roma, Italy}
\address{$^2$ Graduate School of Materials Science, University of Hyogo, Kamigori-cho, Ako-gun, Hyogo 678-1297, Japan}
\address{$^3$ Department of Physics, Kyoto University, Kyoto 606-8502, Japan}
\address{$^4$ Department of Chemistry, Graduate School of Science, Hiroshima University, Higashi-hiroshima, Hiroshima 739-8526, Japan}
\address{$^5$ The Graduate University for Advanced Studies, Institute for Molecular Science, Okazaki, Aichi 444-8585, Japan}
\address{$^6$ JASRI/XFEL, 1-1-1 Kouto, Sayo-cho, Sayo-gun, Hyogo 679-5198, Japan}
\address{$^7$ Physical and Chemical Research Infrastructure Group, Advanced Photon Technology Division, RIKEN SPring-8 Center, 1-1-1 Kouto, Sayo-cho, Sayo-gun, Hyogo 679-5148, Japan}

\ead{adriano.verna@uniroma3.it}

\vspace{10pt}
\begin{indented}
\item[]\today
\end{indented}

\begin{abstract}
An experimental and computational investigation of the space-charge effects occurring in ultrafast photoelectron spectroscopy from the gas phase is presented. The target sample CF$_3$I is excited by ultrashort (100~fs) far-ultraviolet radiation pulses produced by a free-electron laser. The modification of the energy distribution of the photoelectrons, i.e. the shift and broadening of the spectral structures, is monitored as a function of the pulse intensity. The experimental results are compared with computational simulations which employ a Barnes-Hut algorithm to calculate the effect of individual Coulomb forces acting among the particles. In the presented model, a survey spectrum acquired at low radiation fluence is used to determine the initial energy distribution of the electrons after the photoemission event. The spectrum modified by the space-charge effects is then reproduced by $N$-body calculations that simulate the dynamics of the photoelectrons subject to the individual mutual Coulomb repulsion and to the attractive force of the positive ions. The employed numerical method accounts for the space-charge effects on the energy distribution and allows to reproduce the complete photoelectron spectrum and not just a specific photoemission structure. The simulations also provide information on the time evolution of the space-charge effects on the picosecond scale. Differences with the case of photoemission from solid samples are highlighted and discussed. The presented simulation procedure, although it omits the analysis of angular distribution, constitutes an effective simplified model that  allows to predict and account for space-charge effects on the photoelectron energy spectrum in time-resolved photoemission experiments with high-intensity pulsed sources.
\end{abstract}

%
\vspace{2pc}
\noindent{\it Keywords}: Gas-phase photoemission, Space-charge effects, $N$-body simulation, Free-electron lasers, Trifluoroiodomethane

%
%

\section{Introduction}

The extension of the photoelectron spectroscopy (PES) to time-resolved investigations has established itself in recent years as one of the most powerful and promising methods for the study of the electron, spin and lattice dynamics with picosecond and femtosecond time resolution \cite{Bokor_89,Stolow_04,Fadley_10,Schonhense_15}. The pump-and-probe PES technique was successfully employed, for example, in investigating the dynamics of surface chemical reactions \cite{Hockett_11}, demagnetization processes \cite{Rhie_03,Cinchetti_06,Eich_17,Pincelli_19}, charge density waves \cite{Schmitt_08}, image potential states \cite{Pickel_06}, relaxation of photoexcited molecules \cite{Fielding_18}. Advancements with the use of the pump-and-probe technique \cite{Moise_08} are strictly related to the development of stable and intense sources with ultrashort (fs--ps) pulse duration. While the pump signal is often provided by a pulsed optical laser, the list of employed sources for the probe radiation includes the multiplication of a laser fundamental frequency through non-linear crystals \cite{Schmitt_08,Cinchetti_06} or high-harmonic generation (HHG) \cite{Cavalieri_07,Rohwer_11,Cucini_20}, synchrotron light \cite{Glover_03,Pietzsch_07,Pincelli_19}, free-electron lasers (FELs) \cite{Pietzsch_08,Hellmann_12,Oura_14,Squibb_18,Mercurio_19}. This collection of available sources covers a photon-energy range extending from the near ultraviolet to the hard X-rays.

The reduction of the pulse duration to the scale of picoseconds or femtoseconds implies the arrival of a considerable number of photons on the sample within an ultrashort time, with the consequent simultaneous emission of a huge quantity of photoelectrons from a region of space having the size of the beam spot. Photoemitted electrons  are subject to mutual Coulomb repulsion and to the attractive force of the positive charges left in the irradiated material system, determining a variation in their kinetic energy and momentum. The measured  spectrum consequently differs from the genuine distribution of the electron kinetic energy after the photoemission event. The two most important effects that are observed in the photoelectron energy spectrum are the shift of the position and the broadening of the PES structures (valence band or orbitals, core levels, Auger peaks) on the scale of the kinetic energies \cite{Zhou_05}. Space-charge effects on the angular distribution of photoelectrons, i.e. the electron momentum broadening \cite{Passlack_06,Oloff_16,Hellmann_09}, are also of fundamental importance for angle-resolved PES (ARPES) experiments but they will not be investigated in this paper.

Space-charge effects in PES were observed for the first time in the eighties in multiphoton ionization experiments on gases that used optical and near-ultraviolet pulsed lasers \cite{Meek_82,Chupka_85}. Strategies to cancel or reduce them included decreasing the laser intensity \cite{Meek_82,Allendorf_89}, defocusing of the incident beam \cite{Allendorf_89} and the reduction of the gas pressure \cite{Compton_RN_84} but no systematic study or numerical simulation were carried out. Apart the shift and broadening of the PES lines, the observed suppression of lower order peaks in multiphoton ionization processes \cite{Kruit_83,Lompre_85,Hippler_87} was ascribed to the attractive force of the ions that traps the slower electrons and prevents them from reaching the spectrometer \cite{Crance_86a,Crance_86}. Observations in photoemission experiments from solid surfaces followed and more comprehensive studies were carried out using different pulsed sources and exploring a wide range of photon energies \cite{Passlack_06,Faure_12,Frietsch_13,Plotzing_16,Oloff_16,Saule_19, Long_96,Zhou_05,Kuhn_19,Pietzsch_08,Hellmann_10,Hellmann_12a,Fognini_14a, DellAngela_15,Oloff_14}. These investigations evidenced that in PES experiments from solids the space-charge effects are dominated by the numerous low-energy ($\lesssim$20~eV) secondary electrons that constitute the most abundant part of the charge emitted in vacuum. The PES lines are typically shifted towards higher kinetic energies due to the repulsive force of the slow secondary electrons that remain closer to the sample surface and push the faster primary photoelectrons away. Energy shift and broadening are both directly proportional to the number $N_e$ of emitted electrons per pulse (and then to the pulse intensity $I$) and inversely proportional to the linear size $a$ of the beam spot on the sample surface, but  they depend only slightly on the kinetic energy of the PES structures under study \cite{Zhou_05,Hellmann_09}. Space-charge effects were recently studied also in photoemission from liquid solutions \cite{AlObaidi_15}.

While a simple theoretical description of the space-charge limited (SCL) current in thermionic valves or phototubes was well established decades ago \cite{Belinov_09}, a rigorous treatment of the Coulomb interactions in the photoelectron cloud is much more complex but some effective simplified models have been proposed. The space-charge effects are traditionally treated distinguishing between a deterministic contribution, which represents the repulsive force exerted by the average (macroscopic) charge density $\rho(\mathbf{r},t)$ of the electron cloud, and a stochastic contribution, which describes the electron-electron scattering events and takes into account the granularity and the stochastic nature of the electron cloud \cite{Kruit_incoll_97,Schonhense_18}. Generally speaking, the deterministic part leads to a rotation of the particle distribution in the position and momentum space, while the contribution of the stochastic part induces a broadening of the distribution \cite{Schonhense_15a}.
Simulations of the deterministic effects are usually carried out following the trajectories as a function of time of a limited number of particles and apportioning the total charge of the electron cloud among the considered trajectories. The use of advanced softwares like SIMION\textsuperscript{\textregistered}, based on the solution of the Laplace's equation, allowed to include the space-charge effects in the simulation of the motion of the photoelectrons inside a time-of-flight (TOF) spectrometer \cite{Schonhense_18,Greco_16,Schonhense_15}. This approach was particularly effective in the correction of the deterministic effect of the space charge phenomenon in time and angular photoemission from solid samples upon using time-of-flight momentum microscopy \cite{Kutnyakov_20}. In this instrument, a strong electric field is applied in front of the sample, a condition that is not typically encountered in other spectroscopic setup where a field-free region exists close to the sample. Therefore, the large number of slow secondary electrons and the few faster core level and valence electrons emitted from the solid have a different expansion of their momentum-distribution discs in the strong extractor field of the instrument, allowing to reconstruct the experimental spectra limiting the space-charge effect influence.

Long, Itchkawitz and Kabler proposed more than twenty years ago a model to predict the deterministic contribution to the energy broadening of the PES structures. In this model the photoelectrons are described as a negative charge flying between the plates of a spherical capacitor \cite{Long_96}. Despite its numerous simplifications, this model justifies the linear dependence of the energy broadening $\Delta E_B$ on the ratio $N_e/a$ and revealed to be effective in quantifying the order of magnitude of $\Delta E_B$ measured in PES experiments with different pulsed sources \cite{Verna_16}. The stochastic contribution to the space charge has been known for many years in electron microscopy as Boersch effect \cite{Boersch_54} and it induces an energy broadening that adds to the effects of the charge density $\rho(\mathbf{r},t)$ \cite{Passlack_06,Oloff_16,Hellmann_09,Kruit_incoll_97,Schonhense_18}.

The distinction between deterministic and stochastic contributions is overcome by simulation procedures that reproduce the trajectories of as many electrons as those actually photoemitted in the experiment under study. In this way, the simulation takes into account both the long-range effects of the net charge distributions of electrons and ions and the short-range electron-electron interaction. The model presented in this work belongs to the latter class of space-charge simulations. Hellmann~et~al. \cite{Hellmann_09} were the first to propose a $N$-body simulation scheme based on the Barnes-Hut algorithm \cite{Barnes_86}, in which the trajectories of the $N_e$ photoelectrons (representing the photoelectron cloud generated by a single pulse) are determined with an approximate calculation of the force acting on each particle and by performing a leapfrog integration of the equations of motion of each electron. This method requires the use of the Treecode software \cite{Treecode} with slight modifications \cite{Hellmann_09,Verna_16} and has been applied to successfully reproduce the space-charge effects observed in numerous measurements \cite{Hellmann_09,Hellmann_12, Hellmann_12a,Oloff_14,Oloff_16,Oloff_16a} as well as for a feasibility study of future experiments \cite{Verna_16,Greco_16}. Experiments and simulations were also carried out to study the contribution to the space charge from the photoelectrons generated by the optical or near-ultraviolet pump radiation via multiphoton processes and its dependence on the delay time between the pump and probe pulses \cite{Oloff_14,DellAngela_15,Oloff_16,Oloff_16a}.

Nevertheless, the method of a  full-spectrum $N$-body simulation of PES experiments has not been yet explored. The advantage of this method is the possibility to predict or correct for space-charge effects over the entire photoelectron spectrum. Most of the calculations reported in literature for the photoemission from  solid surfaces are focused on studying the space-charge effects on a single specific structure (a core level or the valence band), assuming for the secondary electrons a simple rectangular energy distribution \cite{Hellmann_09,Hellmann_12, Hellmann_12a} or a delta-function distribution \cite{Verna_16,Greco_16} and neglecting the contribution of the higher-energy photoelectrons (other photoemission structures and inelastically scattered electrons). It must be said that a PES experiment from a solid sample is not the easiest test to attempt a full-spectrum simulation. Most of the particles that enter in the simulation are secondary electrons, which constitute the large majority of the overall photoelectrons (80--99~\%) \cite{Henke_81}. The kinetic-energy distribution of these secondary electrons is essentially a very broad bump extending from 0 to about 20~eV, which is expected to be visibly affected by the space charge only at very high pulse intensity. Nevertheless, secondary electrons necessarily occupy the most part of the computation time but they must be included because they are primarily responsible for the distortion of the sharper and more interesting higher energy structures (core levels, valence band and Auger peaks).

A much smaller number of secondary electrons are generated in photoemission from the gas phase. In these experiments the primary photoelectrons themselves are the main source of the shift ad broadening of the PES structures that compose their kinetic-energy distribution, with a small contribution from inelastically scattered electrons. Measurements on gases then offer a simple way to compare the full-spectrum simulations with the experimental results with a lower computational effort with respect to the case of solids. Another simplification for a first test is the use of a moderate photon energy (extreme ultraviolet or soft X-rays) that avoids working on very extended `survey' spectra presenting a lot of different structures.

In the present paper, $N$-body calculations are used to reproduce the space-charge effects in photoemission spectra from gaseous trifluoroiodomethane (CF$_3$I) excited through $h\nu$=95~eV ultrashort ($\sim$100~fs) FEL radiation pulses. The measurements show the evolution of the shift and broadening of the PES structures with increasing the intensity $I$ of the radiation pulses. Spectra acquired at low photon flux and with negligible space charge provide the initial kinetic-energy distribution $Y_{in}(E_K)$ for the photoelectrons in the simulation. The final kinetic-energy distributions $Y_{fin}(E_K)$ obtained from the simulations for different values of $I$ are then compared with the experimental spectra. As we shall see, the good qualitative and quantitative agreement between experiment and simulations guarantees the reliability of the adopted theoretical and numerical approach. The presented computational tool is a first positive step towards the implementation of a standard procedure for taking into account space-charge effects in PES, including investigations on solid samples or using other pulsed sources (first of all HHG and hard X-ray FELs), and provides an instrument for the partial correction of the acquired spectra and for the design of future experiments.

The paper is structured as follows. Section~\ref{sec:experimental} describes the experimental setup and the details of the PES measurements from the CF$_3$I gas. The simulation method and the physical assumptions on which it is based are illustrated in Section~\ref{sec:simulations}. The results of experiment and simulations and the comparison between them are illustrated in Section~\ref{sec:results} while Section~\ref{sec:discussion} discusses the implications of the obtained outcomes. A summary of the work and of the principal conclusions is found in Section~\ref{sec:conclusions}.

\section{Experimental} \label{sec:experimental}

The photoelectron spectroscopy experiment was carried out at beamline-1 \cite{Owada_18} of SACLA XFEL facility \cite{Ishikawa_12} at SPring-8 campus using ultrashort ($\Delta t\sim$100~fs), quasi-monochromatic vacuum ultraviolet (VUV) pulses, the so-called pink-beam, at a repetition rate of 60~Hz. The VUV-FEL photon energy and the average pulse energy were 95~eV and 85~$\mu$J/pulse, respectively. The corresponding photon flux was about $I_0$=5.6$\cdot 10^{12}$ photons/pulse and the beam diameter at the sample position was measured to be about 10~$\mu$m.

\begin{figure}
\centering
\includegraphics[width=12 cm]{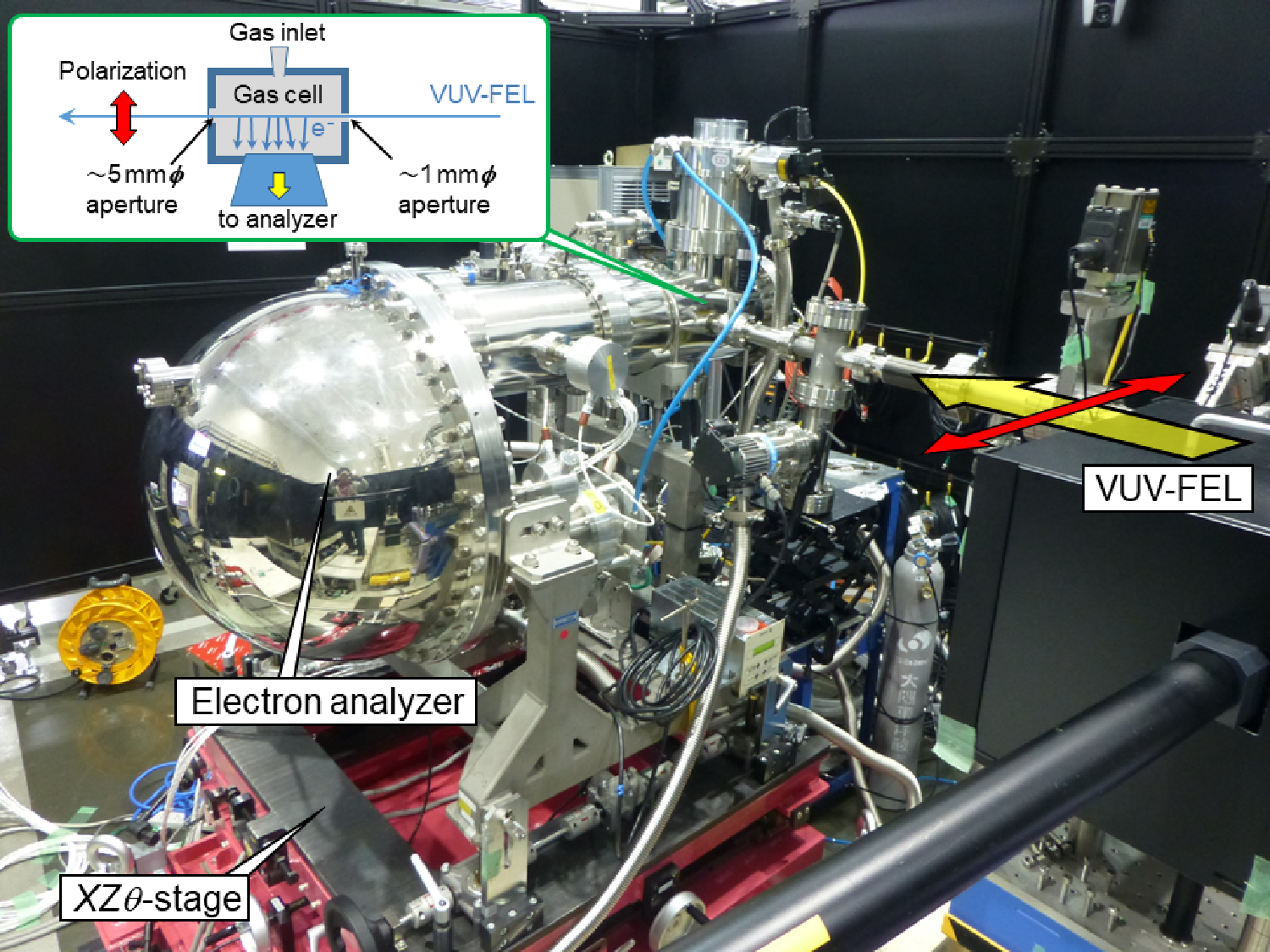}
\caption{Picture of the gas-phase x-ray photoelectron spectroscopy experimental apparatus. The apparatus was mounted on a position-adjustable $XZ\theta$-stage. The VUV-FEL beam comes from the right-hand side (yellow arrow). The inset shows a schematic drawing of the 9~ mm long gas cell which has a 1~mm$\phi$ entrance aperture and a 5~mm$\phi$ exit aperture for the VUV-FEL beam.  In the picture and in the inset the red arrows indicate the polarization direction of the beam.}
\label{fig:setup}
\end{figure}

The experimental setup, equipped with the hemispherical electron energy analyzer SES-2002 and the gas-cell GC-50 (Scienta Omicron \cite{scienta}) used for the measurements of electron spectra is shown in Figure~\ref{fig:setup}. The inset shows a schematic diagram of the 9~mm long gas cell crossed by the VUV-FEL beam ionizing the molecules of the gas. The apparatus is regularly used for the gas-phase x-ray photoelectron spectroscopy experiments at the SPring-8 undulator beamline in the energy range between soft x-ray and hard x-ray \cite{Oura_19}. For the present work, the apparatus was temporarily moved to the SACLA experimental hall. The lens axis of the analyzer was in the horizontal plane at right angles to the VUV-FEL beam direction and parallel to the polarization vector of the incident photons. The apparatus was mounted on a position-adjustable $XZ\theta$-stage, where $X$ stands for the horizontal, $Z$ for the vertical directions and $\theta$ for the rotation, respectively.

All the electron spectra were recorded with the kinetic energy sweep mode of the electron analyzer. The kinetic energy scale of the analyzer was calibrated by measuring the Kr M-NN Auger electron spectrum and comparing the detailed structure observed in 25--62~eV region with the previous work by Aksela et al. \cite{Aksela_H_84}. For the present measurements on CF$_3$I molecule, the pass energy of the hemispherical analyzer was set to 50~eV and a 1.5~mm $\times$ 25~mm analyzer entrance slit (longer side parallel to VUV-FEL beam) was chosen to give a theoretical energy resolution of about 230~meV. The angular acceptance of the analyzer is centered in the polarization direction of the radiation and is strongly elongated in a direction parallel to the beam due to the shape of the entrance slit. The linear magnification of the analyzer optics is $\sim$5. Since beamline-1 of SACLA does not incorporate a monochromator, the bandwidth of the pulsed VUV radiation results larger than 1.0~eV. Measuring the iodine 4d spectrum emitted from the CF$_3$I target at low photon flux, and then with negligible space-charge broadening, an overall spectral energy resolution of about 1.4~eV is estimated. The typical acquisition time for one spectrum was 15~minutes.

In order to study the space-charge effect on the electron spectrum, the pulse intensity at the sample position was adjusted using a 2.6~m-long gas-attenuator chamber filled with N$_2$ gas at a pressure varying between 1 and 90~Pa. Accordingly, the pulse intensity $I$ of the attenuated beam was ranged between approximately 1$\cdot 10^7$ and 5$\cdot10^{12}$ photons/pulse at the sample position. The target sample for the present study was trifluoroiodomethane (CF$_3$I) molecules fed into the main chamber via a gas cell. The pressure in the experimental chamber (outside the gas cell) was maintained at about 1.3$\cdot10^{-3}$~Pa, corresponding to an estimated pressure of 1.26~Pa inside the gas cell. The gas cell was at room temperature during the experiment, hence the density of CF$_3$I in the gas cell was $\rho$=3.0$\cdot 10^{14}$ molecules/cm$^3$. The pulse intensity $I$ can be considered as constant along the length of the gas cell. In fact, the attenuation length of the radiation in the gas cell is given by
\begin{equation*}
\lambda=\frac{1}{\mu}=\frac{1}{\rho\sigma_M}
\end{equation*}
where $\mu$ is the absorption coefficient and $\sigma_M$ is the absorption cross section of a CF$_3$I molecule for $h\nu$=95~eV photons. From the tabulated total atomic photoionization cross sections for C, F, and I \cite{Yeh_85},  an approximated absorption cross section  $\sigma_M$=19.1~Mb can be  estimated. This results in an attenuation length $\lambda$=1.7~m which is much greater than the 10~mm length of the gas cell.

\section{Simulations} \label{sec:simulations}

The present $N$-body simulation of the space-charge effects considers an initial kinetic-energy distribution of the photoelectrons $Y_{in}(E_K)$ defined in the whole range of energies from 0~eV to about $h\nu$. $Y_{in}(E_K)$ represents the photoemission spectrum unaffected by the space charge. The motion of each of the $N_e$ electrons, subject to the repulsion of the other electrons and to the attractive force of the ions, is then calculated until the particles are spaced far enough apart that the Coulomb interactions are negligible. The  final kinetic-energy distribution $Y_{fin}(E_K)$ of the photoelectrons is then compared with the measured spectra, analyzing in detail the modifications induced by the space charge on the various photoemission structures in both experiment and simulation. With this approach, it will be possible to unambiguously verify the reliability of the approximations embedded in the simulation procedure and to control the accuracy of the physical parameters at the base of the calculations, first of all the space-charge-free photoelectron spectrum $Y_{in}(E_K)$.

The calculations were carried out using a modified version of the open-source Treecode software \cite{Treecode} based on the Barnes-Hut algorithm \cite{Barnes_86} that was already applied for the simulation of the space-charge effects in the case of photoemission from a solid surface \cite{Verna_16}. In the simulations, the trajectories of the photoelectrons are calculated considering both their mutual Coulomb repulsion and the Coulomb attraction with the positive ions left behind. The ions can be considered at rest during the flight of the electrons, the root-mean-square (rms) velocity of a CF$_3$I molecule being 196~m/s at 300~K in the ideal gas approximation whereas the speed of an electron with a kinetic energy of just 1~eV is 5.9$\cdot 10^5$~m/s. On the other hand, in the time between two successive pulses (0.017~s) CF$_3$I molecules travel an average distance of about 3~m, indicating that when the new pulse arrives the  ions have left the gas cell or have repeatedly bounced off its walls neutralizing their charge. So it can be assumed that every radiation pulse interacts with a gas of neutral molecules. Photoelectrons are considered as all emitted at the instant $t$=0.

The simulation procedure requires to define the initial configuration of the overall $N_e$ photoelectrons, i.e. their spatial positions $\mathbf{r}_i$ and velocities $\mathbf{v}_i$ at the instant $t$=0 in a defined $O\mathbf{\hat{x}\hat{y}\hat{z}}$ reference system.  The initial position $\mathbf{r}_i$ of each photoelectron $i=1,...,N_e$ is randomly selected inside a cylinder that represents the region of the gas cell crossed by the FEL radiation beam. The radius of the cylinder $a$=5~$\mu$m corresponds to half the measured beam diameter, while the length of the cylinder is $l$=30~$\mu$m and is approximately equal to the spatial extension of the pulse, its duration being $\sim$100~fs as stated above. The axis of the cylinder corresponds to the $\mathbf{\hat{x}}$ coordinate axis. The initial kinetic energy $E_{Ki}$ of each photoelectron is randomly selected according to a normalized probability distribution $f(E_K)$ which essentially reproduces a reference experimental photoelectron spectrum acquired with low radiation intensity ($I=6.6 \cdot 10^7$ photons/pulse at an attenuating N$_2$ gas pressure $p$=80~Pa) and presenting negligible space-charge effects. The transmission function for the spectrometer is assumed determined  by the angular magnification of the entrance lens and hence directly proportional to $(E_K)^{-\frac{1}{2}}$ \cite{Roy_90,Martensson_94}. The low-flux spectrum acquired at $p$=80~Pa is then multiplied by a function proportional to $\sqrt{E_K}$ and a polynomial is used to extrapolate the spectrum in the non-measured low kinetic-energy range 0--7~eV. The area of the obtained function in $E_K$ is finally normalized to 1, giving as a result the probability distribution $f(E_K)$. The modulus of the initial velocity of each photoelectron is then calculated from the initial kinetic energy $E_{Ki}$ as $|\mathbf{v}_i|=\sqrt{2E_{Ki}/m}$. Space-charge effects at different pulse intensities $I$ are then simulated varying the number $N_e$ of photoelectrons emitted from the cylinder at $t$=0. For relatively low values of $N_e$ (less than 30,000), in order to improve statistics the procedure is repeated with different initial configurations and the obtained final kinetic-energy spectra are averaged.

The number of photoemitted electrons $N_e$ that are included in a simulation can be related to the number of photons per pulse $I$ through the ionization cross section. The absorption coefficient $\mu=\rho \sigma_M$ multiplied by $I$ represents the number of absorbed photons per unit length along the beam path, but this does not correspond to the number of photoelectrons per unit length $n_e$ because  Auger electrons are also emitted due to the core-hole decay. The core-hole lifetime is typically of the order of few femtoseconds or less \cite{Sawatzky_incoll_88} and then much shorter than the pulse duration, so that the emission of Auger electrons is assumed as simultaneous to the emission of the primary photoelectrons. Conversely, modifications in the electron energy distribution due to the simulated space-charge effects occur on a timescale at least of the order of hundreds of fs (see below) and then significantly larger than the pulse duration. The 95~eV radiation is above the ionization thresholds for the F~2s and the I~4d core levels, whose binding energies $E_B$ are $\sim$40~eV and 57.8~eV (4d$_{5/2}$) and 59.5 (4d$_{3/2}$)~eV, respectively \cite{Bancroft_84,Yates_86}. For these values of binding energies in the L and N shells the recombination of the core holes can be considered completely non-radiative \cite{Bishop_69,Thompson_book_85}. Previous studies \cite{Bancroft_84,Yates_thesis_86} demonstrated that the contribution of Auger electrons due to the creation of a photohole in the valence orbitals is negligible. So we can assume that the Auger recombination processes originate only from holes in the F~2s (LVV) and I~4d (NVV) states and for the same reasons we also exclude the formation of multiple Auger electrons due to the cascade phenomenon.
The overall number of photoelectrons per unit length of the beam path generated in the gas can then be estimated as the sum of primary photoelectrons and Auger electrons with the formula
\begin{equation}
n_e=\rho\left( \sigma_M+3\sigma_{\mbox{\scriptsize F2s}}+\sigma_{\mbox{\scriptsize I4d}} \right) I
\label{eq:lindensity}
\end{equation}
where $\sigma_{\mbox{\scriptsize F2s}}$=0.60~Mb and $\sigma_{\mbox{\scriptsize I4d}}$=9.1~Mb are the tabulated absorption cross sections for the two core levels at $h\nu$=95~eV \cite{Yeh_85}. The proportionality coefficient $\rho\left( \sigma_M+3\sigma_{\mbox{\scriptsize F2s}}+\sigma_{\mbox{\scriptsize I4d}} \right)$ corresponds to a production of 913 photoelectrons per $\mu$m of beam path and per billion photons.
The number of photoelectrons taken into account in the simulation is then given by $N_e=n_e l$, where $l$ is the length of the cylinder assumed as source of the charged particles. The contribution of the positive ions is represented by a static random distribution of $N_e$ particles with positive charge $+e$ confined in the volume of the cylinder.

For simplicity, all the photoelectrons (photoelectrons from the valence orbitals and from the core levels, Auger electrons, inelastic background) are assumed to be emitted isotropically. This approximation is not too raw considering that, as shown in the low-flux ($I$=$6.6\cdot10^7$ photons/pulse) experimental spectrum (navy-blue continuous line in figure~\ref{fig:experimental}): i) roughly 40\% of the photoelectrons are Auger and low-energy electrons whose angular distribution is not far from isotropic; ii) about 20\% of the photoelectron spectrum are I~4d electrons whose asymmetry factor $\beta$ \cite{Schmidt_book_97} at 95~eV photon energy amounts to 0.8 for an atom \cite{Yeh_book_93} and is lower than 0.5 for the similar CH$_3$I molecule \cite{Lindle_84}; iii) 30\% of the photoelectrons belong to a manifold of molecular valence orbitals of different symmetries and hence with a variety of asymmetry parameters; iv) a mere 10\% of the photoelectrons are F~2s that in atomic approximation are credited of a sharp asymmetry parameter of 2 \cite{Schmidt_book_97}. In order to estimate the effect of anisotropic emission on the shift and broadening of the photoemission structures, we performed a test simulation considering a toy Gaussian initial photoelectron energy distribution $Y_{in}(E_K)$ centered at $E_K$=56~eV (approximately the energy of the F~2s electrons) and with a FWHM of 3.0~eV. The test simulation was carried out comparing the results obtained assuming an asymmetry parameter $\beta$=0 (isotropic emission) or 2 for a density of $n_e$=1970~photoelectron per micron of beam path. This linear density of photoelectrons corresponds in the simulation of the CF$_3$I spectrum to an intensity $I$=2.2$\cdot 10^9$ photons/pulse through equation (\ref{eq:lindensity}). For isotropic emission, the obtained final energy distribution $Y_{fin}(E_K)$ has a maximum at 52.5$\pm$0.3~eV and a FWHM of 11.6$\pm 0.3$~eV whereas for the anisotropic distribution the maximum of $Y_{fin}(E_K)$ is at 52.2$\pm$0.3~eV and its FWHM is 13.6$\pm$0.3~eV. The isotropic assumption causes a negligible error on the energy shift and an inaccuracy of 15\% for the energy broadening for the case of the most anisotropic emission, which concerns a limited number of the overall photoelectrons. Identical results were found in the test simulations also for a lower value of the linear density of photoelectrons, namely $n_e$=465~photoelectrons/$\mu$m ($I$=5.1$\cdot 10^8$ photons/pulse).

Given the initial positions and velocities of the photoelectrons at instant $t$=0, the simulation code calculates the force acting on each electron and then the new position and velocity after a leapfrog integration time step $\delta t$ \cite{Hellmann_09,Verna_16,Treecode}. The procedure is then iterated simulating the dynamics of the photoelectron cloud in the interval between $t$=0 and a chosen ending time $t_f$. It is then possible to obtain a histogram of the kinetic energy of photoelectrons at different times $t$ and the final histogram at $t=t_f$ is the simulated spectrum $Y_{fin}(E_K)$ that shall be compared with the experiment.

The integration time step must be carefully chosen because too large values determine a poor time resolution and produce results that significantly differ from the real integrated solutions; on the other end, a too short time step is uselessly time consuming \cite{Verna_16}. Moreover, in the presented case a time step shorter than the pulse duration ($\sim$0.1~ps) makes the approximation of instantaneous emission of the photoelectrons meaningless. Simulation tests carried out with different time steps $\delta t$ revealed that negligible variations in the results are found decreasing $\delta t$ below 0.125~ps. Furthermore, this value of integration time step is slightly longer than the pulse duration but shorter than the time scale at which significative modifications in the simulated electron energy distribution are observed (about 1~ps as demonstrated below) and is then adopted for the calculations.

\section{Results} \label{sec:results}

\begin{figure}
\centering
\includegraphics[width=12 cm]{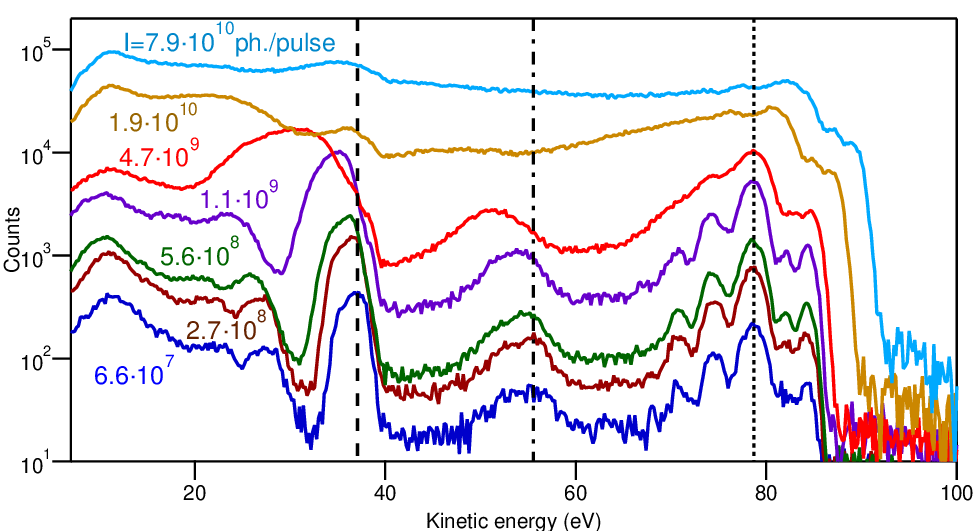}
\caption{Experimental photoelectron spectra for different pulse intensity $I$ from the CF$_3$I gas in the kinetic-energy range 7--100~eV using a FEL photon energy of 95~eV. The labels next to the displayed curves indicate the corresponding intensity in terms of number of photons per pulse. The vertical lines indicate the energy positions at low photon flux for the maxima of the I~4d core level (37.1~eV), of the F~2s core level (55.6 ~eV) and of the most intense component of the valence band, the 3\emph{e} state \cite{Yates_86} (78.7 eV), and make it easier to evidence the shift of the corresponding structures.}
\label{fig:experimental}
\end{figure}

In figure~\ref{fig:experimental} we show a selection of the experimental photoelectron spectra acquired in the kinetic-energy range 7--100~eV for different values of the exciting pulse intensity $I$. In the spectra reported in the figure, the pulse intensity varies between 6.6$\cdot 10^7$ and 7.9$\cdot 10^{10}$ photons per pulse and these values of flux were obtained changing the attenuating N$_2$ gas pressure in the range 80--30~Pa. It is evident a progressive shift towards lower kinetic energy of most of the structures, in particular the I~4d ($\sim$37~eV) and the F~2s ($\sim$56~eV) core levels. This can be ascribed to the attractive force of the positive ions left in the irradiated cylinder. Nevertheless, at low photon flux no shift is present for the structure of the valence orbitals  that extends in the 65--85~eV kinetic-energy range. This absence of energy shift is caused by the shielding effect of the lower-energy electrons that during the flight are mostly located between the ``motionless'' positive ions and the faster photoelectrons from the valence band. Only for very high photon flux, the valence band experiences a clear shift towards {\em higher} kinetic energies. The maximal kinetic-energy value of the valence band, which is associated to photoelectrons from the highest occupied molecular orbital (HOMO) structure, increases from $\sim$85~eV to 91~eV in the reported spectra and exceeds 100~eV at full pulse intensity (not shown in figure~\ref{fig:experimental}). The experimental measurements also manifest an energy broadening of the reported structures. For example, it is evident that the detailed structure of the valence band \cite{Bancroft_84,Yates_86} is lost when the intensity reaches 4.7$\cdot 10^9$~photons/pulse.

\begin{figure}
\centering
\includegraphics[width=12 cm]{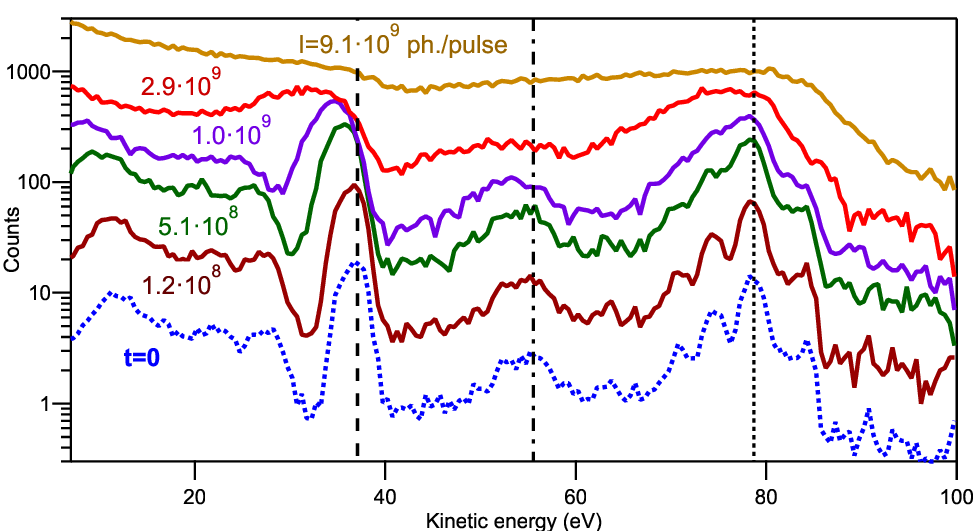}
\caption{Simulated photoelectron spectra for different values of the pulse intensity $I$ (1.2$\cdot 10^8$--9.1$\cdot 10^9$ photons per pulse), indicated by the labels next to each displayed curve. The displayed kinetic-energy distributions are calculated at time $t$=125~ps after the arrival of the FEL pulse. The lowest curve (blue dotted line) represents the initial kinetic-energy distribution of photoelectrons at instant $t$=0.}
\label{fig:simulation}
\end{figure}

The photoelectron spectra obtained from the $N$-body simulations are shown in figure~\ref{fig:simulation}. The simulations were carried out considering different values of linear density of photoelectrons $n_e$ in the range 0.11--8.4$\cdot 10^3$ electrons/$\mu$m/pulse and the distortions induced by the space charge understandably become more pronounced with increasing $n_e$. The values of the pulse intensity $I$ reported next to every spectrum is obtained from $n_e$ through equation (\ref{eq:lindensity}), ranging approximately from 1$\cdot 10^8$ to 1$\cdot 10^{10}$ photons/pulse. The reported spectra are the histograms of the kinetic energy of the photoelectrons at a time $t$=125~ps after the arrival of the radiation pulse and the photoemission event. Extending the simulation time $t$ beyond 125~ps does not cause further changes in the energy distribution of the photoelectrons. Notably, similarly to the experimental spectra, the simulations predict a shift towards lower kinetic energy for the I~4d and of the F~2s core levels and a fixed position for the valence-band states that move towards higher kinetic energy only at very large photon fluxes. Simulations also foresee an energy broadening of the photoemission structures.

It is important to compare quantitatively the values of energy shift and broadening and their dependence on the pulse intensity as obtained from the experiment and from the simulation procedures. To this end,  we focus on the two core-level peaks I~4d and F~2s. The position in terms of kinetic energy of their maximum and their energy width are extracted from both the experimental and the simulated spectra through a fitting procedure. The profile of both these peaks was fitted as the sum of two Gaussian functions with the additional contribution of a simple linear background. In particular, the use of a double Gaussian is necessary for the I~4d peak because of the spin-orbit splitting  between d$_{3/2}$ and d$_{5/2}$ components, that results 1.76~eV in agreement with the tabulated data \cite{xray_booklet}, with a fixed branching ratio of 0.67.  The presence of a doublet structure in the F~2s peak is instead ascribed to a symmetry breaking induced by the photoionization process which splits the molecular orbitals into states of $e$ and $a_1$ character \cite{Banna_75}. Analyzing the experimental spectra acquired with lower intensity $I$ (small space-charge effects), the separation between the two components is found to be 3.4~eV while the ratio between the intensities results 0.63 with the main peak at higher kinetic energy. These values are very similar to those obtained on the analogous CF$_3$H molecular gas using $h\nu$=132.3~eV radiation \cite{Banna_75}.

\begin{figure}
\centering
\includegraphics[width=14.0 cm]{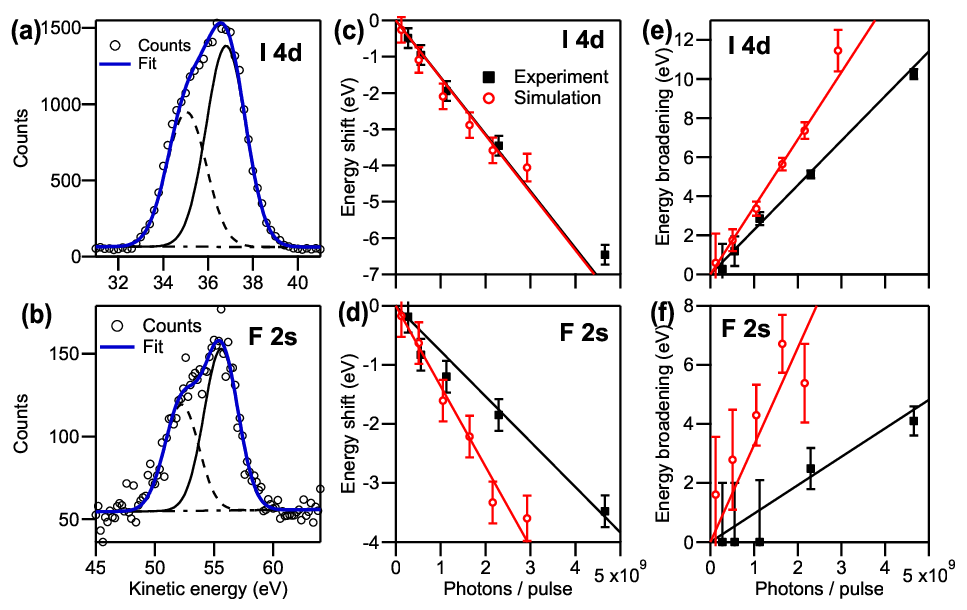}
\caption{\emph{(a)-(b)}: I~4d and F~2s photoemission peaks measured with intensity $I$=2.7$\cdot$10$^8$ photons/pulse (empty black circles) and their fitting (blue solid line) given by the sum of two Gaussian functions (dashed and solid black lines) and of a linear background (dash-dotted black line). \emph{(c)-(f)}: Energy shift $\Delta E_S$, \emph{(c)} and \emph{(d)}, and energy broadening $\Delta E_B$, \emph{(e)} and \emph{(f)}, calculated for the experimental (full black squares) and simulated (empty red circles) I~4d and F~2s photoemission peaks as a function of the number of photons per pulse $I$. For the simulated spectra the number of photons per pulse $I$ is estimated from the number of photoelectrons per $\mu$m and per pulse $n_e$ using equation (\ref{eq:lindensity}). The straight lines are linear fitting to the reported points according to the relations $\Delta E_S=aI$ and $\Delta E_B=bI$, the intercept being fixed at zero.}
\label{fig:quantities}
\end{figure}

\begin{table}
\caption{\label{tab:slopes} Slopes of the best-fitting straight lines for the measured and simulated energy shift $\Delta E_S$ and broadening $\Delta E_B$ of the I~4d and F~2s peaks as a function of the number of photons per pulse $I$ (figure~\ref{fig:quantities}~(c)--(f)). $a$ is the slope of $\Delta E_S$ and $b$ is the slope of $\Delta E_B$ and both the coefficients are expressed in electronvolt per billion photons in a single pulse.}
\begin{indented}
\item[]\begin{tabular}{@{}lllcll}
\br
            & \multicolumn{2}{c}{\textbf{I 4d}}                      && \multicolumn{2}{c}{\textbf{F 2s}} \\
            & $a$ (eV/$10^9$ ph.)   & $b$ (eV/$10^9$ ph.)   && $a$ (eV/$10^9$ ph.)   & $b$ (eV/$10^9$ ph.) \\
            \cline{2-3} \cline{5-6} \noalign{\smallskip}
Experiment  & $-1.570\pm 0.032$      & $2.284\pm 0.038$     && $-0.767\pm 0.032$     & $0.964\pm 0.051$ \\[0.08 cm]
Simulation  & $-1.583\pm 0.088$     & $3.45\;\: \pm 0.12$        && $-1.363 \pm 0.088$    & $3.29\;\: \pm 0.48$ \\
\br
\end{tabular}
\end{indented}
\end{table}

The fitting of the experimental I~4d and F~2s photoemission lines acquired with a moderate intensity $I$=2.7$\cdot$10$^8$ photons/pulse are reported in figures~\ref{fig:quantities}~(a) and (b), respectively. The significant quantities that are extracted from the fitting are the energy position of the main peak $E_M$ (the $j$=5/2 component for the I~4d core level and the state with $e$ character for F~2s) and the full width at half maximum (FWHM) of the overall photoemission line $w$. The energy shift of the peak can then be calculated as the difference $\Delta E_S=E_M-E_{M0}$ and the energy broadening as the root of the difference of squares $\Delta E_B=\sqrt{w^2-w_0^2}$ \cite{Hellmann_09}, where $E_{M0}$ and $w_0$ are the main-peak position and the FWHM without space charge. $E_{M0}$ and $w_0$ can be obtained from the analysis of the simulated spectra at the initial time $t$=0 and they results for I~4d peak $E_{M0}$=37.3$\pm$0.2~eV and $w_0$=3.3$\pm$0.2~eV while for the F~2s peak they are $E_{M0}$=55.8$\pm$0.2~eV and $w_0$=6.0$\pm$0.4~eV. The found values of energy shift $\Delta E_S$ as a function of the number of photon per pulse $I$ are reported in figures~\ref{fig:quantities}~(c) and (d) for the I~4d and F~2s peaks, respectively. The obtained values of energy broadening $\Delta E_B$ of the two peaks are presented in figures~\ref{fig:quantities}~(e) and (f). The four panels (c)--(f) in figure~\ref{fig:quantities} allow to compare the effects of space charge in the experimental (full black squares) and simulated (empty red circles) spectra. Both experiments and simulations show that the energy shift and broadening have an approximately linear dependence on the pulse intensity (i.e. the number of photoelectrons), a behaviour already observed in photoemission from solid samples in previous experimental and theoretical investigations \cite{Long_96,Zhou_05,Hellmann_09,Verna_16}. The straight lines in the aforesaid panels are linear fitting to the reported experimental and simulated values and their slopes, reported in table~\ref{tab:slopes}, are the coefficients of proportionality between the space-charge effects ($\Delta E_S$, $\Delta E_B$) and the pulse intensity $I$, expressed  in terms of electronvolts per billion photons in a single pulse. It is evident that, for the same pulse intensity, simulation in general tends to slightly overestimate the space-charge effects with respect to the results provided by the experiment. Nevertheless, the difference between experiment and calculation in the obtained values of slope reported in table~\ref{tab:slopes} is never greater than a factor $\sim$3, a discrepancy which is similar to that found in analogous studies for photoemission from solid surfaces \cite{Hellmann_09}. This corresponds to an acceptable quantitative agreement between experiment and simulations, considering the simplicity of the implemented model and the critical approximations that have been adopted.

\begin{figure}
\centering
\includegraphics[width=12.0 cm]{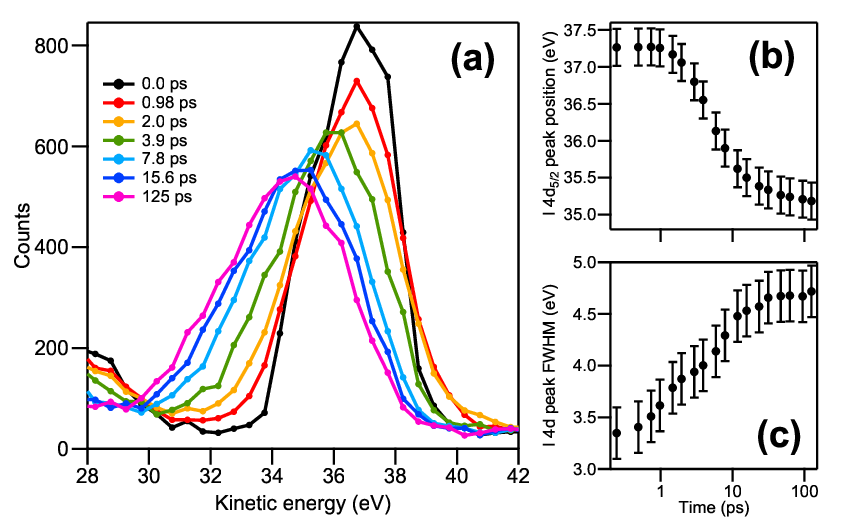}
\caption{\emph{(a)}: Simulated I~4d core-level photoemission peak at different times $t$ with a linear density of photoelectrons $n_e$=0.96$\cdot$10$^3$~electrons/$\mu$m/pulse, that corresponds to an intensity $I$=1.0$\cdot$10$^9$~photons/pulse according to equation~(\ref{eq:lindensity}). \emph{(b)-(c)}: The energy position $E_M$ of the maximum of the I~4d$_{5/2}$ component \emph{(b)} and the FWHM of the overall I~4d peak \emph{(c)} extracted from the fitting of the simulated spectra in panel~(a) and reported as a function of time $t$.}
\label{fig:time}
\end{figure}

Using the simulation procedure it is also possible to study the dynamics of the energy distribution of the photoelectrons. As an example, in figure~\ref{fig:time}~(a) we report the evolution of the simulated I~4d spectrum for a moderate radiation intensity ($I$=1.0$\cdot$10$^9$~photons/pulse) in the time range $t$=0--125~ps from the arrival of the radiation pulse and the concomitant photoemission event. The values of the energy position $E_M$ and of the FWHM $w$ as a function of time extracted from the fitting of the photoemission peaks are reported in the panels (b) and (c). From the data reported in these figures, it is evident that the two space-charge effects present a clearly different dynamics. A progressive broadening of the electron energy distribution  starts immediately after the photoemission event while the maximum of the peak initially does not move.  The shift of the peak towards lower kinetic energy begins to be evident at a time $t\approx$4~ps. The variations of the I~4d energy position and FWHM mostly occur in the first 10~ps and the evolution of the photoemission structure can be considered completed at the end of the simulation at $t$=125~ps.

\section{Discussion} \label{sec:discussion}

The progressive shift towards \emph{lower} kinetic energies of the PES structures with increasing the pulse intensity, observed in the experimental spectra and reproduced by the simulations, constitutes the most striking difference with the case of photoemission from solid surface, where a shift towards \emph{higher} kinetic energies is commonly experienced. In the case of photoemission for the gas phase, the reduced kinetic energy of the photoelectrons is clearly ascribed to the attractive force exerted by the positive ions which basically remains in the cylinder of gas crossed by the FEL radiation beam during the flight of the negative particles. The positive charge of the ions exerts on the photoelectrons, which are escaping and moving away from the irradiated gas region, a force directed back towards the cylinder, determining a decrease of their kinetic energy.

In photoemission from solids, the space-charge effects are instead dominated by the secondary electrons in the 0--20~eV kinetic-energy range \cite{Zhou_05,Verna_16}, which, as already stated, constitute the majority of the overall photoelectrons \cite{Henke_81}. The secondary electrons and the higher-energy electrons of the other spectral structures (valence band, core levels, Auger electrons and inelastic background) emerge at the same time from the radiation spot on the sample surface but are quickly spatially separated on account of their different velocities. The numerous and slower secondary electrons remain closer to the surface and as a result of the  Coulomb repulsion push away the other electrons, which in this case acquire a higher kinetic energy. Positive charges are present also in photoemission from solids, in the form of mirror charge of the photoelectrons if the surface is metallic whereas for an insulator the positive ions left in the region of the radiation spot cannot be neutralized. Nevertheless, the attractive force that the positive charge on the sample surface exerts on the high-energy photoelectrons is mostly shielded by the interposed secondary electrons, resulting on the whole in an increased kinetic energy \cite{Zhou_05,Hellmann_09}. This screening effect of the positive charge is absent in the photoemission from a rarefied gas because the production of secondary electrons is much smaller, due to the low density and the extremely larger electron mean free path.

In gas-phase photoemission, the effect of the attractive force of the positive ions and the related negative shift of the PES structures depends on the kinetic energy. Photoelectrons having different kinetic energy (and consequently different velocity) are subject to a progressive spatial separation after their common emission from the irradiated cylinder of gas. Electrons emitted from the valence band have the highest kinetic energy (70--85~eV) and move away from the cylinder faster than the other photoelectrons. The attractive force that the positive ions exert on these faster photoelectrons is mostly shielded by the interposed photoelectrons with lower kinetic energy that oppose a counteracting repulsion. This explains the absence of energy shift for the valence band structures at moderate intensities revealed in the experiment (figure~\ref{fig:experimental}) and reproduced by the simulation procedures (see figure~\ref{fig:simulation}).

Immediately after the photoemission event there is no spatial separation between positive ions, slower electrons and faster electrons, all the charged particles being contained in the irradiated cylinder. Once the photoelectrons have left the cylinder, the positive charge of the almost immobile ions exerts on the photoelectrons around them an attractive force directed back towards the cylinder. The work performed by this attractive force reduces the overall kinetic energy of the photoelectrons determining the negative shift. On the contrary, random Coulomb interactions and scattering events of a photoelectron with the other electrons and the ions may either increase or decrease its net kinetic energy. This is the stochastic contribution to the space-charge effects, already mentioned in the Introduction, and is an important source of the energy broadening of the PES structures. This explains why the simulated evolution of the I~4d peak (see figure~\ref{fig:time}) indicates that the effect of energy broadening temporally precedes the energy shift. A random modification in the kinetic energy of the photoelectrons and the resulting energy broadening starts immediately after the photoemission event, when all the charged particles are in the irradiated cylinder. The energy shift requires a spatial separation between ions and electrons and can occur only when most of the electrons have left the positively charged cylinder. For example, in the simulation of the evolution of the  I~4d peak, the energy shift appears only about $t$=4~ps after the emission, a time during which the I~4d photoelectrons ($\sim$36~eV) travel a distance of $\sim$14~$\mu$m. Considering a isotropic emission of the photoelectrons, this corresponds to a mean distance traveled in a direction perpendicular to the cylinder's axis of $\sim$11~$\mu$m, which is comparable with the diameter of the beam. This indicates that the negative energy shift of the I~4d photoelectrons starts only when most of them have left the cylinder of the irradiated gas.

\section{Conclusions} \label{sec:conclusions}
The present study illustrates an experimental and computational investigation of the space-charge effects on the energy distribution of electrons photoemitted from a molecular gas, using the extreme-ultraviolet radiation generated by a FEL apparatus as a pulsed source. Shift and broadening of the PES structures were monitored as a function of the pulse intensity $I$. In contrast to photoemission from solid samples, where the PES structures are subject to Coulomb interactions that increases their kinetic energy, in photoemission from gases a progressive shift towards lower kinetic energies is observed. This different behaviour is ascribed to the secondary electrons that dominate the space-charge effects in photoemission from solids but are much less numerous in the case of gases and cannot shield the attractive force of the positive ions. Nevertheless, the extent of this negative energy shift depends on the kinetic energy and is almost absent for the photoelectrons originated from the valence orbitals due to the shielding effect of the lower-energy electrons. Only for very high photon fluxes the valence photoemission structures are shifted, but towards higher kinetic energies. All the evidenced space-charge effects are well reproduced by the $N$-body simulations that take into account the Coulomb interactions among the photoelectrons and with the positive ions. In the new presented approach, a reliable initial energy distribution for all the photoelectron is provided by a low-intensity measurement and the simulation procedure calculates the modifications occurring in the full photoelectron spectrum. Despite some strong approximations adopted to reproduce the investigated system, a reasonable quantitative agreement between the experimental and simulated results was obtained. It must be borne in mind that the low energy resolution of the experiment ($\sim$1.4~eV) limits the level of confidence attributable to this model and that future experiments with better energy resolution will be a tougher test for the effectiveness of the proposed model. Simulations also give information on the time evolution of the space-charge effects and show that the broadening effect precedes the energy shift of the photoemission structures. The results of the present study on photoemission from the gas phase validate the effectiveness of the proposed simulation procedure, which can be extended to experiments on solid surfaces and to the use of other types of pulsed sources. The illustrated method can be also applied on molecular dynamics calculations based on the particle-to-particle (p2p) approach, where the contribution of the positive charge of the ions is of critical importance.

\section*{Acknowledgements}
The experiments at BL1 of SACLA were carried out with the approval of JASRI (proposal nos.~2017A8070 and 2017B8021). The authors thank the operating and engineering staff of SACLA for their support during the experiments. We are grateful to the members of the engineering team of the RIKEN SPring-8 Center for their technical assistance.

Financial support from PRIN~2015  NEWLI: NEW LIght on transient states in condensed matter by advanced photon-electron spectroscopies, Protocol:  2015CL3APH, is greatly acknowledged.


\section*{References}


\begin{thebibliography}{10}
\bibitem{scienta}
The analyzer as well as the gas-cell were originally supplied by Gammadata
  Scienta\texttrademark{}. Present company name is
  ScientaOmicron\texttrademark{}.
\end{thebibliography}

\providecommand{\newblock}{}

\end{document}